\documentclass[11pt]{article}

\usepackage[T1]{fontenc}
\usepackage{newtxtext,newtxmath}
\usepackage{microtype}
\usepackage{geometry}
\usepackage{setspace}
\usepackage{graphicx}
\usepackage{caption}
\usepackage{mathtools}
\usepackage{bm}
\usepackage{cite}

\usepackage{amsthm}
\usepackage{siunitx}
\usepackage{cite}
\usepackage{booktabs}
\usepackage{listings}
\usepackage{xcolor}
\usepackage{algorithm}
\usepackage{algpseudocode}
\definecolor{codegray}{rgb}{0.4,0.4,0.4}
\definecolor{lightgray}{rgb}{0.95,0.95,0.95}

\lstdefinestyle{heyland}{
    language=Python,
    basicstyle=\ttfamily\footnotesize,       
    numbers=left,                     
    numberstyle=\tiny\color{codegray},
    stepnumber=1,
    numbersep=8pt,
    frame=single,                     
    rulecolor=\color{lightgray},
    framerule=0.5pt,
    backgroundcolor=\color{lightgray},
    keywordstyle=\bfseries,           
    stringstyle=\ttfamily,            
    commentstyle=\itshape\color{codegray}, 
    breaklines=true,                  
    breakatwhitespace=true,
    columns=fullflexible,
    tabsize=4,
    columns=flexible,
    keepspaces=true,
    showstringspaces=false,
}

\usepackage{hyperref}
\hypersetup{
    colorlinks=true,
    linkcolor=blue,
    urlcolor=blue,
    citecolor=blue,
    pdfborder={0 0 0}
}
\makeatletter
\def\Hy@raisedlink#1{#1}  
\makeatother

\theoremstyle{plain}

\theoremstyle{definition}

\title{\Large\bfseries
HeylandCircle: A Computational Framework for the Geometric Reconstruction of the Heyland Circle Diagram
}

\let\oldeqref\eqref
\renewcommand{\eqref}[1]{Eq.~\oldeqref{#1}}

\author{
\large Anubhav Gupta\thanks{Email: \texttt{anubhav.gupta@colorado.edu}}\;
\href{https://orcid.org/0000-0002-3216-868X}{\includegraphics[height=10pt]{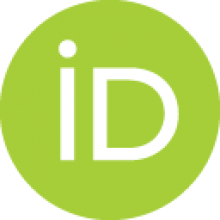}}\\[4pt]
\small University of Colorado Boulder, CO 80303, USA\\
\small In Orbit Aerospace Inc., Torrance, CA 90501, USA
\and
\large Abhinav Gupta\thanks{Email: \texttt{abhinav.gupta@colorado.edu}}\;
\href{https://orcid.org/0009-0006-8863-1457}{\includegraphics[height=10pt]{Orcid-ID.png}}\\[4pt]
\small University of Colorado, Boulder, CO 80303, USA
}

\date{}

\begin{document}
\maketitle
\vspace{-0.75em}

\begin{abstract}
The Heyland circle diagram is a classical graphical tool for representing the steady–state behavior of induction machines using no–load and blocked–rotor test data. While widely used in alternating–current machinery texts, the diagram is typically presented as a hand–constructed aid and lacks a standardized computational formulation. This paper presents \texttt{HeylandCircle}, a computational framework that reconstructs the classical Heyland circle diagram directly from standard test parameters. The framework formalizes the traditional geometric construction as a deterministic, reproducible sequence of geometric operations, establishing a clear mapping between measured data, fixed geometric objects, and steady–state operating points. Quantities such as power factor, slip, output power, torque, and efficiency are obtained through explicit geometric relationships on the constructed diagram. Validation using a representative textbook example demonstrates close agreement with classical results. The framework provides a computational realization of the traditional Heyland diagram suitable for instruction, analysis, and systematic extension.
\end{abstract}

\noindent\textit{Keywords:}
induction machines; Heyland circle diagram; geometric methods; computational geometry; Python framework

\section{Introduction}
\label{sec:intro}
The Heyland circle diagram is a classical geometric representation of the steady–state behavior of induction machines \cite{heyland1906graphical}. Constructed using only no–load and blocked–rotor test data, the diagram compactly encodes the variation of input current, slip, torque, and power factor across operating conditions \cite{langsdorf1937theory}. Its primary appeal is pedagogical: a single geometric construction exposes relationships that are less transparent in purely algebraic equivalent–circuit formulations.

Despite its instructional value, the Heyland diagram remains largely a hand–constructed tool and is seldom incorporated into modern computational workflows due to the inherent subjectivity and lack of precision associated with graphical methods. As a result, no widely adopted specification exists for its reproducible construction and use within contemporary analysis pipelines.

This work introduces \texttt{HeylandCircle}, a reference computational framework that reconstructs the classical Heyland circle diagram directly from standard steady–state test parameters. The framework formalizes the traditional geometric construction as a deterministic and reproducible computational pipeline. The formulation establishes a consistent mapping between measured test data, fixed geometric objects defining the diagram, and steady–state performance quantities extracted from these objects through explicit geometric relationships. While the circle diagram is occasionally attributed to Ossanna or Behrend in specific regional contexts, we adopt the Heyland nomenclature in this framework, as it remains the standard reference in the global literature on induction machine tests.

The objective is to render classical pedagogy precise and extensible. The diagram is preserved in its traditional geometric form, while its construction is expressed in a manner suitable for scripting, validation, and integration into modern computational workflows. Formal analytic and geometric justifications of the construction—including existence, uniqueness, and equivalence to the steady–state equivalent–circuit current locus—are developed separately in \cite{gupta2025mobius}, allowing the present work to focus exclusively on computational specification rather than proof.

Within this context, \texttt{HeylandCircle} serves as a stable computational reference for the classical diagram and establishes a baseline for systematic generalizations. In particular, departures from ideal circularity—arising from non–ideal effects or parameter dependencies—naturally motivate extensions such as elliptical loci, which are addressed in subsequent work.

The authors have previously explored a numerical reconstruction of the circle diagram in an earlier textbook \cite{gupta2012testing} using spreadsheet–based computations and plotting tools. While pedagogically effective, that implementation did not isolate the geometric rules of the classical construction in a reusable computational form. The present work builds upon that experience by providing a unified, computationally defined formulation of the Heyland diagram suitable for use as a reference computational layer. Following classical machine–theory conventions, the active (wattful) component of current is plotted along the vertical axis, corresponding to the real component of the complex current phasor.

The remainder of the paper is organized as follows. Section~\ref{sec:classical} fixes the geometric objects defining the classical diagram. Section~\ref{sec:framework} describes the computational formulation and construction sequence. Section~\ref{sec:geometric} presents the geometric interpretation of steady–state performance quantities. Section~\ref{sec:validation} verifies the framework against a standard textbook example, and Section~\ref{sec:applications} discusses integration into broader computational workflows. Section~\ref{sec:conclusion} summarizes the scope and role of the framework.

\section{The Classical Heyland Circle Diagram}
\label{sec:classical}

\subsection{Role of the Classical Diagram}
Introduced as a geometric construction derived from experimental test results, the diagram connects measured data, equivalent–circuit interpretation, and performance estimation within a single graphical object. Although modern analysis typically emphasizes numerical equivalent–circuit solvers or time–domain simulation, the Heyland diagram remains distinctive in that it encodes the complete steady–state operating envelope through a fixed geometric locus.

In the present work, the classical diagram is treated as a geometric object to be specified and reconstructed computationally. This section therefore fixes the inputs, geometric primitives, and relationships that define the diagram independently of drafting conventions.

\subsection{Test Data and Geometric Anchors}
The classical Heyland diagram is defined by two standard steady–state tests \cite{fitzgerald2002electric, gupta2012testing}:

\begin{itemize}
\item the no–load test, providing the magnitude and phase angle of the no–load current $(I_0, \phi_0)$, and
\item the blocked–rotor test, providing the magnitude and phase angle of the blocked–rotor current $(I_{sc}, \phi_{sc})$ measured at a reduced voltage and referred to rated conditions.
\end{itemize}

These measurements are represented as complex current phasors and define two fixed points in the complex current plane. All subsequent geometric elements of the diagram are constructed from these anchors together with a reference horizontal through the no–load point.

The line joining the no–load and blocked–rotor phasors defines the \emph{output line}. The center of the Heyland circle is obtained as the unique point on the reference horizontal lying on the perpendicular bisector of this line, and the circle radius is defined as the distance from this center to either test point. Auxiliary geometric objects—including the torque chord, slip line, and efficiency line—are constructed deterministically from these primitives.

This specification fixes the diagram independently of scale, drafting conventions, or manual construction choices and provides a consistent basis for computational implementation.

\subsection{Analytic Context}
Under standard assumptions of balanced, sinusoidal steady–state operation, the per–phase equivalent circuit predicts that the stator current of an induction machine traces a circular locus in the complex plane as slip varies. Formal analytic and geometric justifications of this result are developed in \cite{gupta2025mobius}. The present work adopts these results as a theoretical foundation; the focus here is on specifying a reproducible computational construction of the classical diagram based on the resulting geometric definitions.

\section{The \texttt{HeylandCircle} Computational Framework}
\label{sec:framework}
This section describes the computational formulation used to reconstruct the classical Heyland circle diagram from standard steady–state test data. The formulation, referred to as \texttt{HeylandCircle}, provides a deterministic mapping between experimentally obtained current phasors and the geometric objects defining the classical diagram.

The framework adopts the geometric definitions fixed in Section~\ref{sec:classical} and expresses the classical construction as a sequence of explicit computational operations. All diagram elements arise directly from prescribed geometry; no analytical re–derivation, numerical optimization, or curve fitting is introduced. The objective is to render the classical construction reproducible and computationally accessible.

\subsection{Design Considerations}
The formulation is guided by three principles:

\begin{itemize}
\item \textbf{Geometric correspondence}: each computed object maps directly to a named element of the classical diagram;
\item \textbf{Determinism}: identical test data produce identical constructions and performance quantities;
\item \textbf{Structural clarity}: the construction is decomposed into elementary geometric operations suitable for reuse or extension.
\end{itemize}

Together, these principles ensure that the framework functions as a reference specification rather than an implementation–specific visualization routine.

\subsection{Input Data Representation}
The framework accepts the standard steady–state measurements:
\[
(I_0,\phi_0), \quad (I_{sc},\phi_{sc}), \quad
V_{\mathrm{rated}}, \quad V_{sc}.
\]

These quantities define the no–load and blocked–rotor current phasors referred to rated voltage and populate a minimal data structure. No additional machine parameters are required. Optional scaling factors may be applied for visualization or unit conversion but do not affect the underlying geometry.

\subsection{Computational Abstraction of the Classical Construction}
Given the input phasors, the Heyland diagram is reconstructed by evaluating a fixed set of geometric primitives—lines, perpendiculars, and intersections—defined relative to the no–load and blocked–rotor current points. The computational framework operates directly on these primitives rather than reproducing manual drafting steps. For completeness, the traditional hand–construction procedure is summarized separately in Appendix~A.

\subsection{Rendering and Usage Illustration}
For visualization and practical use, the computed geometric objects may be rendered as an annotated circle diagram. Rendered elements include the current locus, output line, torque chord, and labeled geometric points. Rendering does not modify the underlying geometric construction.

\begin{lstlisting}[style=heyland]
from heylandcircle import CircleDiagram, MachineTestData

data = MachineTestData(
    I0=6,               # No-load current (A)
    phi0_deg=85,        # No-load phase angle (deg)
    Isc=12,             # Blocked-rotor current (A)
    phi_sc_deg=69.0667, # Blocked-rotor phase angle (deg)
    V_rated=400,        # Rated voltage (V)
    V_sc=100,           # Blocked-rotor test voltage (V)
    P_rated_kw=5.6,     # Rated power (kW)
)

cd = CircleDiagram(data, use_full_circle=True)
fig, ax = cd.plot()
\end{lstlisting}

\begin{figure}[ht!]
    \centering
    \includegraphics[width=0.99\linewidth]{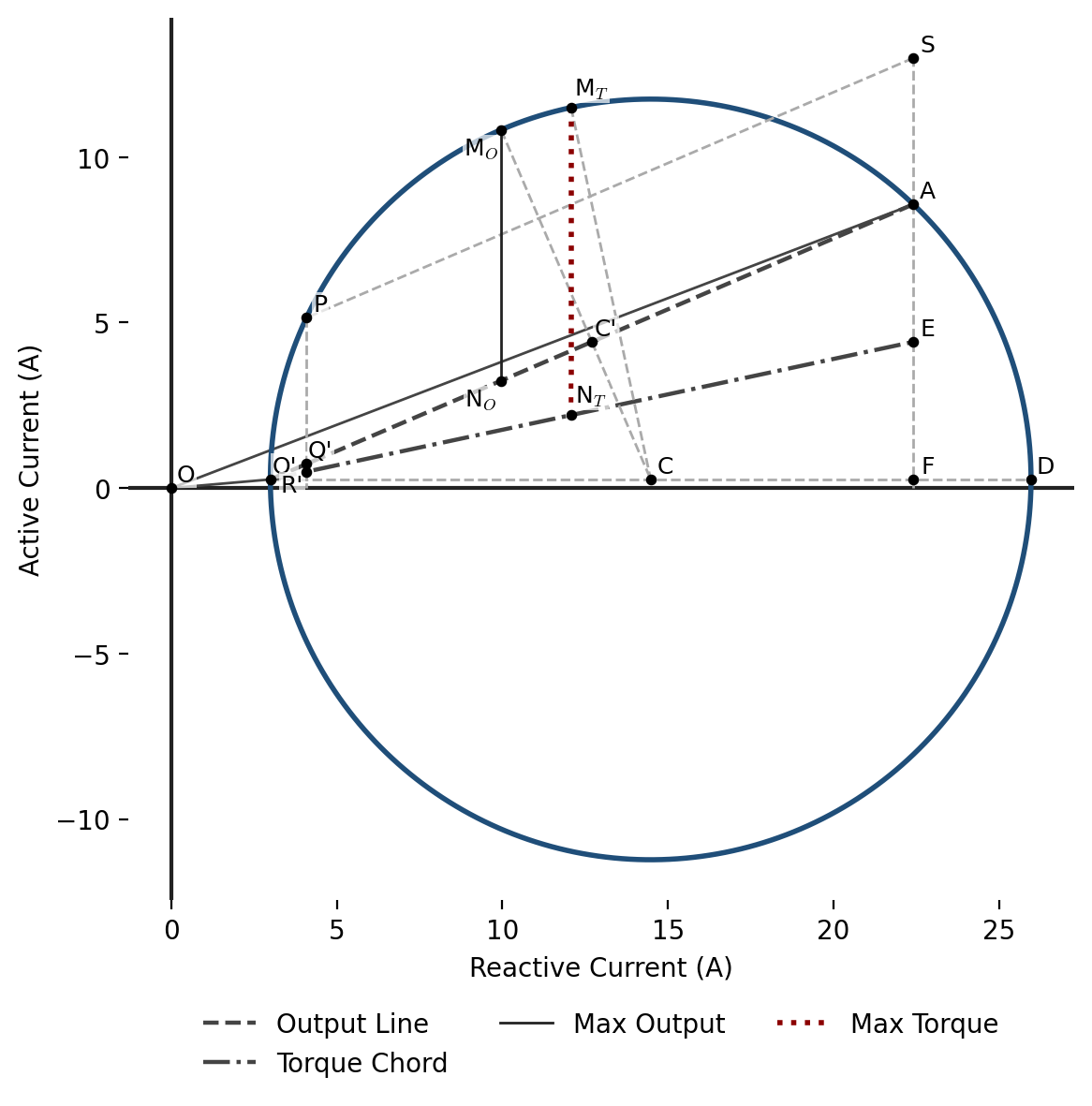}
    \caption{Annotated Heyland circle diagram illustrating slip and
             performance reference constructions.}
    \label{fig:full-circle}
\end{figure}

Optional configuration flags control diagram scaling, axis normalization, and inclusion of generating regions; these parameters affect visualization only and do not modify the underlying geometric construction. Figure~\ref{fig:full-circle} shows the full current locus produced by the computational construction, illustrating both motoring and generating regions of operation.

\section{Geometric Interpretation of Steady--State Performance Quantities}
\label{sec:geometric}
Once the geometric construction of the Heyland circle diagram is fixed, steady--state performance quantities may be interpreted as queries on the resulting geometric object once the diagram has been calibrated. The diagram is treated as a current locus together with a set of reference lines and projection rules, from which operating quantities are obtained directly through geometric relationships. No analytical evaluation of equivalent--circuit expressions is required at this stage.

\subsection{Current Phasor and Power Factor}
Each operating condition is represented by a point on the current locus corresponding to the stator current phasor. The in--phase (wattful) component of current is measured along the vertical axis of the diagram, while the quadrature component lies along the horizontal axis. The input current magnitude and power factor are therefore encoded geometrically by the position and orientation of the operating point relative to these reference axes.

\subsection{Slip and Torque Mapping}
Slip is not treated as an explicit coordinate of the diagram, but is associated with each operating point through calibrated geometric constructions. In the classical Heyland formulation, slip may be obtained as a ratio of vertical intercepts measured between the operating point and designated reference lines. Alternatively, slip may be read from a calibrated slip scale constructed along the slip reference line. Both constructions are purely geometric and yield consistent values within the expected tolerance of graphical methods, and are therefore treated as equivalent for the purposes of this work.

Electromagnetic torque is interpreted geometrically through projections onto the torque chord. The maximum--torque operating point is identified as the point at which this chord is tangent to the current locus, providing a purely geometric criterion independent of circuit parameter evaluation.

\subsection{Operating Regimes and Efficiency Interpretation}
The Heyland diagram also permits graphical interpretation of operating regime and efficiency through auxiliary reference constructions. Relative intercepts and projections on the diagram provide a geometric visualization of efficiency variation across operating conditions in the traditional manner. Operating points lying beyond the zero--slip reference correspond to negative slip and indicate generating operation, which is identified directly from the geometry of the constructed locus without additional analytical conditions.

\section{Validation}
\label{sec:validation}
Validation is performed using the standard worked example of the Heyland circle diagram presented in Theraja and Theraja \cite{theraja2014textbook}, Chapter 35, Example 35.7, with parameters taken directly from the published test data in the reference. The objective is to verify faithful reproduction of classical Heyland–diagram results, rather than to assess physical model fidelity.

\begin{table}[h]
\centering
\caption{Comparison between textbook values and \texttt{HeylandCircle} outputs.}
\begin{tabular}{lccc}
\toprule
\textbf{Quantity} & \textbf{Textbook} & \textbf{Computed} & \textbf{Error (\%)} \\
\midrule
Maximum output (kW)    & 10.80   & 10.62   & --1.66 \\
Maximum torque (Sync kW)    & 14.07   & 13.01   & --7.53 \\
Power factor & 0.80 & 0.78 & --2.50 \\
Efficiency (\%)          & 88.04  & 85.86  & --2.47 \\
Slip (\%)           & 4.7 & 5.01 & 6.59 \\
\bottomrule
\end{tabular}
\end{table}

The textbook efficiency value is not explicitly tabulated in the reference example, but is inferred from the published Heyland circle diagram using the standard graphical efficiency construction. Percentage error is reported as the relative difference between the computed value and the corresponding textbook value, normalized by the textbook value.

The observed deviations are consistent with the tolerance of classical graphical constructions and primarily reflect differences between exact geometric intersections and approximate straight-line readings commonly employed in manual drafting. Slip values are particularly sensitive to drafting conventions and reference-intercept placement and are therefore expected to exhibit larger variation than current, torque, or efficiency estimates.

Additional test cases produce consistent agreement and are omitted here for brevity, as the objective is verification of construction fidelity rather than statistical evaluation.

\section{Applications and Extensions}
\label{sec:applications}
The computational formulation presented in this paper supports applications in analysis, instruction, and method development. As a computational realization of the classical Heyland diagram, the framework provides a modern companion to traditional hand--drawn constructions used in classical instruction and analysis. Parameter variations that influence the shape and position of the current locus—such as changes in leakage reactance, magnetizing reactance, or rotor resistance—may be explored directly through their geometric effect on the diagram, enabling rapid and intuitive examination of steady--state behavior while preserving consistency with classical textbook interpretations \cite{carpaneto2002teaching}.

From an analytical perspective, the framework serves as a compact reference for steady--state performance predicted by the equivalent circuit. Because the Heyland diagram encodes the complete steady--state operating envelope, the computational reconstruction may be used as a baseline against which results from time--domain simulation or finite--element analysis can be compared. Deviations from the circular locus highlight the influence of non--ideal effects, including magnetic saturation, deep--bar rotor behavior, and frequency--dependent parameter variation, which lie outside the assumptions of the classical construction.

The explicit computational specification also provides a foundation for systematic extensions of the classical diagram. When the assumptions leading to circularity are relaxed—most notably through slip--dependent or frequency--dependent impedances—the current locus departs from a circle. Such departures motivate generalized geometric constructions, including elliptical loci, which are addressed separately in subsequent work. By fixing the classical construction as a reference case, the present framework enables these extensions to be formulated and compared in a consistent manner.

Because all geometric objects and performance quantities are computed explicitly, the formulation integrates naturally into Python--based analysis workflows. It may be used as a validation tool, as a front--end for parameter studies, or as an interpretable steady--state complement to higher--fidelity numerical models.

\section{Conclusions}
\label{sec:conclusion}
This paper presented a reference computational formulation of the classical Heyland circle diagram derived from standard induction–machine test data. By expressing the traditional geometric construction as a deterministic sequence of computational operations, the framework fixes the diagram’s geometric objects and performance interpretations in a reproducible and implementation–independent manner.

The contribution of this work lies in formalizing the classical diagram as a computational object suitable for validation and systematic extension. By fixing the geometric construction independently of drafting conventions, the framework establishes a stable reference for steady–state analysis and future generalizations. An interactive parameter–sweep interface has been developed to explore sensitivity to test parameters; this functionality lies beyond the scope of the present paper.

\appendix

\section*{Appendix A: Classical Geometric Construction Procedure}
This appendix summarizes the traditional hand--drawn procedure for constructing the Heyland circle diagram as described in classical textbooks. The algorithm is included for reference and comparison only; the computational framework presented in the main text adopts fixed geometric definitions rather than reproducing these drafting steps explicitly.

\begin{algorithm*}[ht]
\caption{Classical Heyland Circle Construction}
\begin{algorithmic}[1]
\State Establish reference axes $Oa$ (horizontal) and $Ob$ (vertical).
\State Plot the no--load current vector $I_0$; denote its tip by $O'$.
\State Draw a line through $O'$ parallel to the horizontal axis.
\State Plot the blocked--rotor current vector $I_{sc}$; denote its tip by $A$.
\State Join $O'$ and $A$ to form the output line.
\State Construct the perpendicular bisector of $O'A$ and extend it to
       intersect the horizontal through $O'$ at point $C$.
\State With center $C$ and radius $O'C$, draw the Heyland circle.
\State From point $A$, draw perpendiculars to establish torque and slip
       reference constructions.
\State Construct the torque line, slip line, and auxiliary reference
       projections as required.
\end{algorithmic}
\end{algorithm*}

\bibliographystyle{ieeetr}
\bibliography{references}

\end{document}